\documentclass[amsmath,amssymb,amsbsy,aps,prb,twocolumn,superscriptaddress,10pt]{revtex4-2}

\usepackage{graphicx}
\usepackage{wasysym}
\usepackage{color}

\newcommand{\DTO}{Dy$_\text{2}$Ti$_\text{2}$O$_\text{7}$\ }
\newcommand{\HTO}{Ho$_\text{2}$Ti$_\text{2}$O$_\text{7}$\ }
\newcommand{\K}{\text{K}}
\newcommand{\mK}{\text{mK}}

\begin{document}

\title{Investigation into the low temperature state of the spin-ice material \DTO }
	 
\author{Mariano Marziali Bermúdez}
\email{mmarziali@df.uba.ar}
\affiliation{Instituto de Física de Buenos Aires (IFIBA), UBA-CONICET, Argentina.}
\affiliation{Departamento de Física, Facultad de Ciencias Exactas y Naturales, Universidad de Buenos Aires, Argentina.}

\author{R. A. Borzi}
\affiliation{Instituto de Física de Líquidos y Sistemas Biológicos (IFLYSIB), UNLP-CONICET, Argentina.} 
\affiliation{Departamento de Física, Facultad de Ciencias Exactas, Universidad Nacional de La Plata, Argentina.}

\author{D. A. Tennant}
\affiliation{Department of Physics and Astronomy, University of Tennessee, Knoxville, TN 37996, USA}
\affiliation{Department of Materials Science and Engineering, University of Tennessee, Knoxville, TN 37996, USA}

\author{S. A. Grigera}
\affiliation{Instituto de Física de Líquidos y Sistemas Biológicos (IFLYSIB), UNLP-CONICET, Argentina.} 
\affiliation{Departamento de Física, Facultad de Ciencias Exactas, Universidad Nacional de La Plata, Argentina.}
	
\date{\today}

\begin{abstract}
The thermal equilibrium properties of the spin-ice material \DTO, including specific heat, magnetization, and spin correlations, could be successfully reproduced by a model featuring magnetic interactions up to the third nearest neighbor and long-ranged dipolar forces.  With the best-fit parameters, the model predicts an ordered ground state which breaks the cubic symmetry of the lattice.  In this work, we analyze results from a neutron scattering experiment in which, instead of sharp Bragg peaks, a diffuse pattern was observed down to $300~\mK$, despite very slow cooling [A. M. Samarakoon et al., Physical Review Research 4,033159 (2022).].  Using a reverse Monte Carlo approach, we found compatible spin configurations, analyze the suitability of antiferromagnetic spin chains as building blocks for the ground-state and provide various measures of correlation and calculate their energy. Our analysis suggests that while infinitely long chains are not present in the experimental configuration, antiferromagnetic spin chains provide a good approximation of the data. There are indications of possible evidence for short-range chains, but further investigation is needed for confirmation.
\end{abstract}
	
\maketitle

\section{Introduction}

Magnetism is not only interesting in its own right, but is also a fascinating field to study complex phenomena. It provides a framework where simple models --often realizable in real-life materials to varying degrees of approximation-- enable the exploration of a diverse array of phenomena arising from many-body interactions \cite{moessner2006geometrical,starykh2015unusual,lhotel2020fragmentation,knolle2019field}. These range from simple ordered states to correlated spin liquids with emergent quasiparticle excitations such as Majorana fermions \cite{kitaev2006anyons} or magnetic monopoles in spin ice \cite{Castelnovo2008,ryzhkin2005magnetic}, whose collective behavior is a matter of current research \cite{ross2011quantum,castelnovo2012spin,slobinsky2021monopole}. 

Spin-ice is a class of materials that has been the focus of recent theoretical and experimental activity \cite{udagawa2021spin}. In these magnets, the frustrated spins mimic the disorder observed in the hydrogen atoms in water ice. Spin-ice materials are typically pyrochlore lattices with magnetic moments located at their vertices.  \DTO\cite{ramirez1999} and \HTO\cite{Harris1997}  are two prototypical spin-ice materials. Both feature spins with a large magnetic moment ($\mu \sim 10~\mu_B$) constrained to lie along the local $[111]$ axis joining both neighboring tetrahedra.  If interactions are restricted to nearest-neighbor ferromagnetic interactions, energy is minimized by satisfying Bernal and Fowler's \textit{ice rule} \cite{Bernal1933}: out of the four spins at the vertices of each tetrahedron, two should point inwards and two outwards. This leads to the eponymous characteristic of spin ices, the formation of an exponentially degenerate ground state characterized by Pauling's entropy. 

Inspired by the classical experiment by Giauque and Stout on water ice \cite{giauque1936entropy}, Ramirez et al. \cite{ramirez1999} measured the specific heat of \DTO (DTO)   and determined the residual entropy of this material to be close to Pauling's estimate. This established DTO as a spin-ice material, free from the significant nuclear-spin contributions that obscured some results in \HTO , 
and also raised fundamental questions regarding the role of the dominant dipolar interaction present in these materials \cite{siddharthan1999ising,den2000dipolar,bramwell2001spin,siddharthan2001spin,melko2001long,shastry2003spin}. In particular, how is it that the NN model entropy is stable to the inclusion of a dominant long-range interaction? The answer was elegantly given by Isakov, Moessner and Sondhi \cite{isakov2005spin}: the ice-rule on the Ising pyrochlore gives rise to an emergent gauge structure that manifests itself in dipolar correlations, and thus the general structure of the spin-ice NN model is preserved by the addition of dipolar interactions.  At sufficiently low temperatures the small but non-zero differences between these two models lead to long-range order at $T \approx 0.13D$ as found in loop Monte Carlo simulations by Melko, den Hertog, and Gingras (MDG) \cite{melko2001long}.  This MDG state belongs to the spin-ice manifold and exhibits zero magnetization. It can be described as an antiferromagnetic arrangement of ferromagnetic spin chains, oriented along axes rotated by 
$\pi/4$ relative to the original lattice \cite{melko2001long}.

Additional terms are needed in the spin-ice Hamiltonian to properly account for the different experimental observations, such as magnetic susceptibility under applied field, heat capacity, diffuse neutron scattering, and an effort was made to determine an empirical Hamiltonian from the experimental data by including exchange interactions up to two different kinds of third nearest neighbors \cite{Fennell2004,ruff2005finite,yavors2008dy,Borzi2016,samarakoon2020machine}.  Some of these terms preserve the ordered ground state, but others, such as the difference between the two third-nearest-neighbor exchange constants, result in different long-range ordered states \cite{McClarty2015,Borzi2016}. The common factor among these states is the existence of spin chains.  The low-energy minimum depends on very small differences in exchange interactions and is surrounded by a multitude of competing states lying at energies that are only marginally higher \cite{McClarty2015}.

The experimental situation has not been as clear-cut as its theoretical counterpart.  No long-range order has been determined in a spin ice material; as recognized in early work \cite{snyder2001spin}, the dynamics of spin ice materials becomes extremely slow at temperatures much higher than the expected onset of long-range order \cite{snyder2001spin,Matsuhira2002,clancy2009,quilliam2011dynamics,Matsuhira2011,yaraskavitch2012spin,eyvazov2018common,wang2021,samarakoon2022anomalous}.  In 2013, D. Pomaranski and colleagues \cite{Pomaranski2013} reported specific heat measurements taken over very long relaxation times.  The data showed an upturn at low temperatures that reduced the residual entropy below Pauling's value and was suggestive of the onset of a long-range ordered state, albeit at much higher temperatures than what was to be expected from the theoretical models. Further works, both theoretical and experimental \cite{henelius2016refrustration,giblin2018pauling}, attribute this upturn to the presence of random disorder in the samples. 

Recently, some of us have investigated the low-temperature properties of isotopic DTO, which is perhaps the cleanest example of a spin-ice material. Noise measurements and neutron scattering experiments using different cooling protocols were combined with computer simulations and theoretical modeling \cite{samarakoon2022anomalous,samarakoon2022structural,hallen2022dynamical}. This work reveals a picture of glassiness without disorder---a form of magnetic structural spin glass characterized by intrinsic anomalous low-temperature dynamics.  The magnetic excitations, subject to topological and microscopic restrictions, are constrained to move in dynamical fractal patterns, evident in the anomalous noise spectrum, that make the relaxation time grow faster than an Arrhenius law.  This picture is in agreement with previous experimental results \cite{snyder2001spin,Matsuhira2002,clancy2009,quilliam2011dynamics,Matsuhira2011,yaraskavitch2012spin,wang2021,eyvazov2018common}, and implies that no long-range order will be detected within the time frames accessible to normal experimental conditions. Recent neutron scattering experiments with cooling protocols for months are in agreement with this observation \cite{giblin2018pauling,samarakoon2022structural}.

The central question driving this work is whether it is possible to identify {\em precursors} to long-range order following an extended cooling process. More specifically, the aim is to determine whether evidence of spin chains---the fundamental building blocks of the low-temperature states predicted by theory \cite{melko2001long,McClarty2015,henelius2016refrustration,Borzi2016}---can be observed. With this objective in mind,  we analyze neutron diffuse scattering using a reverse Monte Carlo (RMC) approach.  RMC has been used in the past to analyze neutron scattering in frustrated magnets \cite{gardner2011slow,paddison2012empirical}.  In the particular case of spin-ice, RMC was used on powder averaged data in \textit{stuffed} spin-ice to study the distribution of couplings \cite{gardner2011slow}, and was also able to recover the ice rule from 1D powder scattering function input \cite{paddison2012empirical}. Here we revisit the DTO neutron scattering data published in ref. \onlinecite{samarakoon2022structural} and analyze the slow-cooled 2D diffuse scattering pattern obtained on an isotopic single crystal to find traces of chain ordering. 
	
\section{Methods}
\subsection{Sample and Neutron Scattering}
\begin{figure}[bt]
\centerline{\includegraphics[width=\columnwidth]{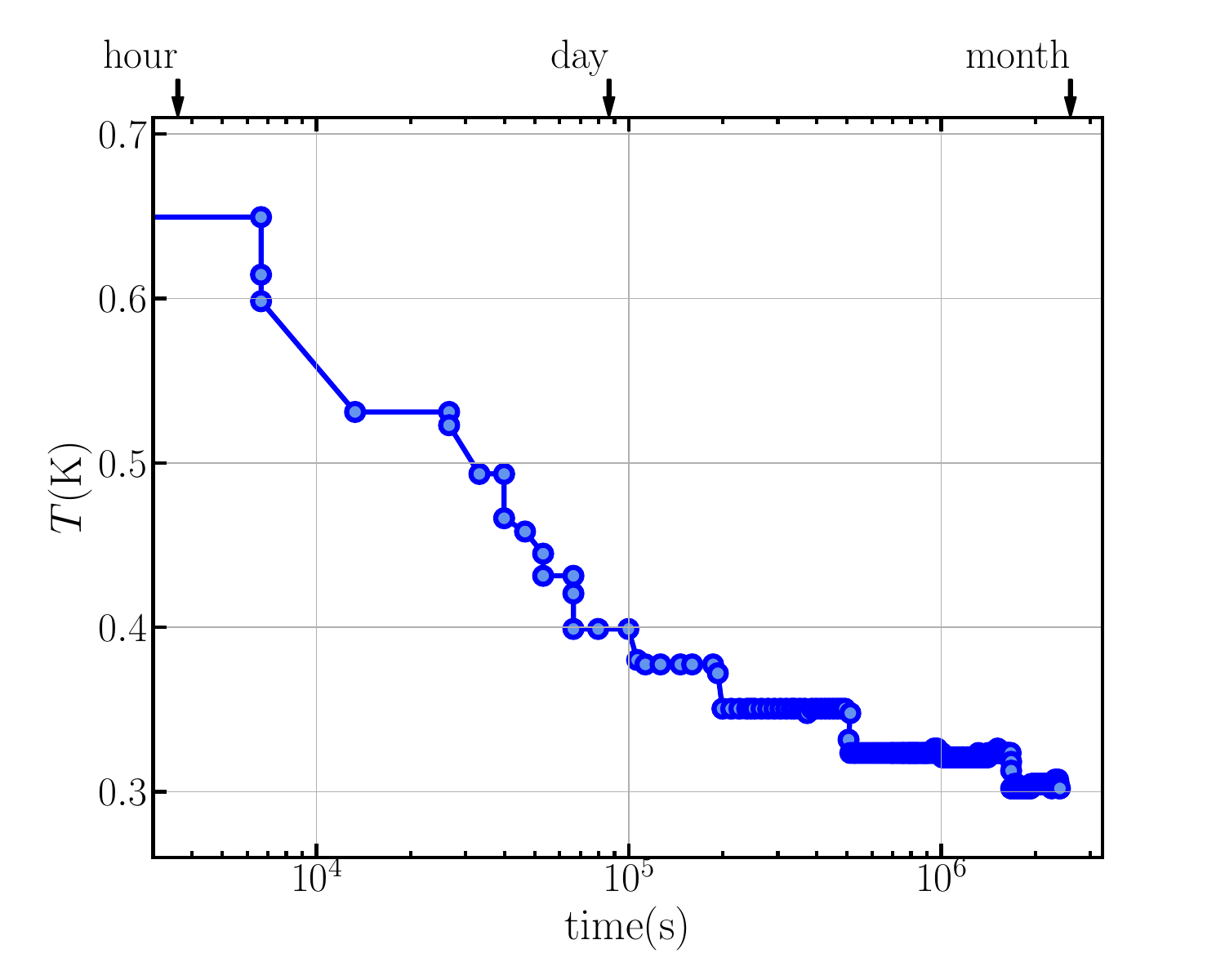}}    
\caption{{\em Cool down procedure of Ref. \onlinecite{samarakoon2022structural}.} Temperature recorded by a thermometer attached at the base of the sample function of time on a semi-logarithmic scale in seconds. For reference, the upper axis shows the span of an hour, a day and a month. The temperature starts at 0.65~K, where the sample reaches thermal equilibrium within seconds, and it is lowered towards the target temperature of 0.3~K.}    
\label{fig:cooling}
\end{figure}
The data we are analyzing in this work, first published in ref. \onlinecite{samarakoon2022structural}, were obtained in an experiment conducted on a 95--98$\%$  $^{162}\text{Dy}$-enriched \DTO single crystal grown using an optical floating-zone method \cite{li2013single}.  The crystal was cut into a cylinder of approximately 10~mm diameter and 50~mm length along the $\langle 110 \rangle$ direction with a mass $\approx 200~{\rm mg}$.  The neutron scattering experiments were made using the E2 flat cone diffractometer at the BER II research reactor, Helmholtz-Zentrum Berlin (see \cite{samarakoon2022structural} for details). Diffuse scattering was measured with four Denex $30 \times 30 ~\mathrm{cm}^2$ position-sensitive detectors. The single-crystalline sample was oriented so as to measure $S(\mathbf{q})$ across the $(h,h,l)$ plane in reciprocal space. To reduce background contamination of the diffuse scattering, a measurement at saturated field was subtracted, replacing Bragg peaks in the background with a suitable region of scattering in $\mathbf{q}$ space. This procedure allows an accurate and reliable determination of the magnetic scattering.
 
\begin{figure}[bt]
\centerline{\includegraphics[width=\columnwidth]{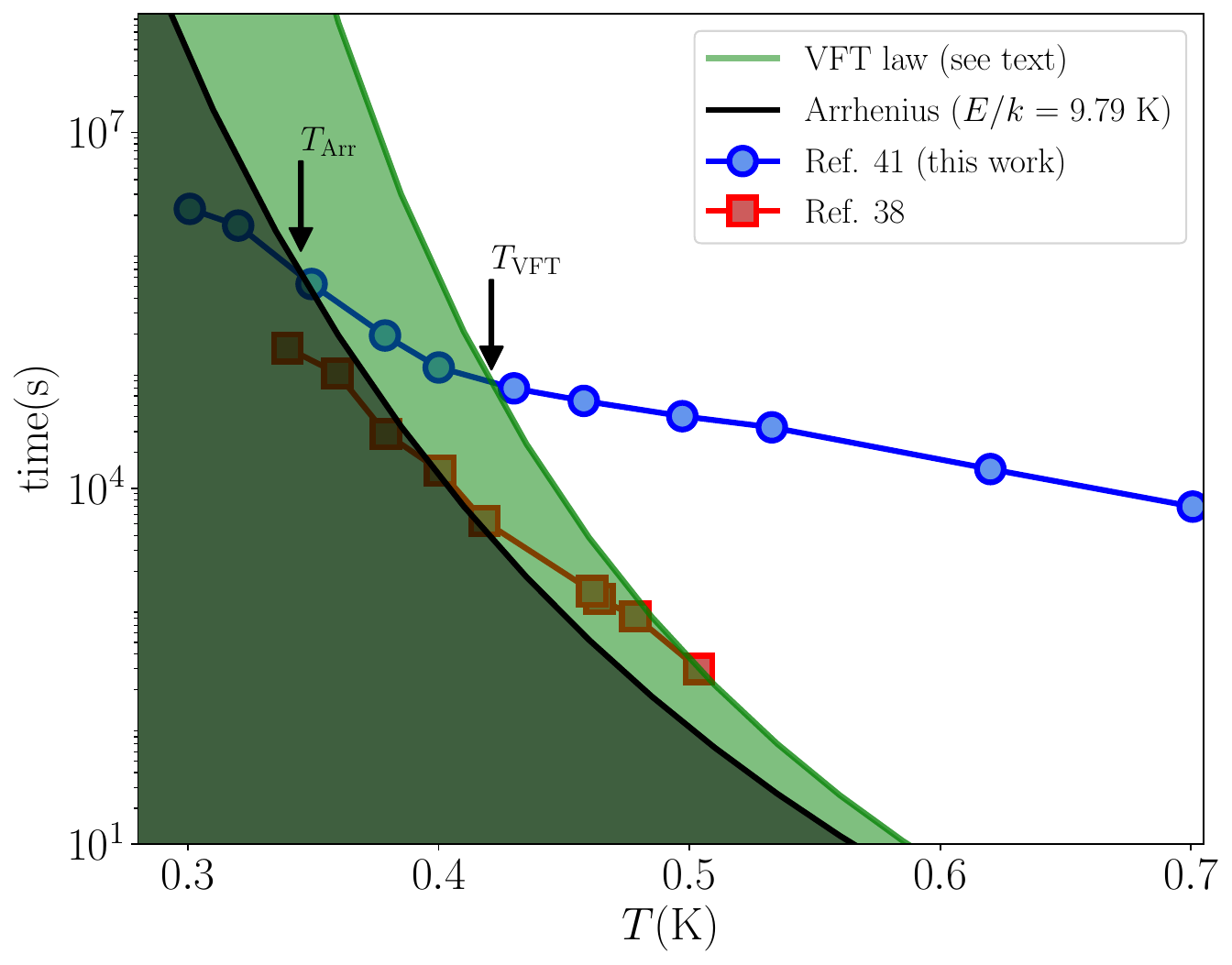}}
\caption{{\em Equilibration times.} Comparison of the time spent at each temperature during the cool-down procedure of the neutron scattering experiment from ref.  \onlinecite{samarakoon2022structural} analyzed in this work (blue curve) with the expected characteristic times of spin-ice, as derived from two different experimental estimates. The first estimate follows an Arrhenius law (dark green) and the second a Volger-Fulcher law (light green).  When the curves lie above the shaded parts of the figure the system is expected to be in equilibrium.  We label $T_{\rm Arr}$ and $T_{\rm VFT}$ the two law-dependent estimates of the lowest equilibrium temperature achieved.  The red squares correspond to the specific heat experiment of ref. \onlinecite{Pomaranski2013}. }
\label{fig:relax}
\end{figure}

The slow cooling procedure used in the experiment is shown in Fig. \ref{fig:cooling}. The temperature ($T$) was gradually lowered during approximately 27 days towards a target temperature of 300~mK. Measurements were taken at this target temperature. The objective of this cool-down procedure was to achieve equilibrium at the lowest possible temperature. This is achieved when the time spent at a given temperature $T$ is longer than the characteristic time. In the case of spin-ice, the former is not unequivocally determined.  Pomaranski et al. \cite{Pomaranski2013} followed a cool-down protocol based on an Arrhenius law.  Recent work \cite{samarakoon2022structural} indicates that a closer description of the experimental situation is a Vogel-Fulcher-Tammann (VFT) law common for glasses, $\tau(T) = \tau_0 \exp\left[A/(T-T_{\rm VF})\right]$, with a faster growth of the characteristic time as temperature is lowered.  Fig. \ref{fig:relax} shows the time spent at each temperature during the cooldown procedure of the specific heat experiment of ref. \onlinecite{Pomaranski2013} (red curve) and the neutron scattering experiment of ref. \onlinecite{samarakoon2022structural} (blue curve) compared with both the expectation for the characteristic time based on an Arrhenius law (dark green), with $E/k = 9.79 \K$ and a VFT law (light green) with $\tau_0 = 7.0(3) \times 10^{-6}~\text{s}$, $T_{\rm VF} = 0.18(2)~\K$, and $A = 6.0(9)~\K$ (see \onlinecite{samarakoon2022structural}).   When the curves lie above the shaded parts of the figure the system is expected to be in equilibrium.  We label $T_{\rm Arr}$ and $T_{\rm VFT}$ the two law-dependent estimates of the lowest equilibrium temperatures achieved, corresponding to the intercept of the cool-down curve for each case.  In the case of an Arrhenius relaxation, the lowest equilibrium temperature of this procedure corresponds to $T_{\rm Arr}=0.345(10)~\K$, while in the VFT case $T_{\rm VFT}=0.42(1)~\K$.  

\subsection{Reverse Monte Carlo approach}

In order to determine spin configurations compatible with the experimental results, we apply a reverse Monte Carlo approach.

\subsubsection{Acceptance criterion}

We explored the spin configuration space in a $16 \times 16 \times 16$ conventional pyrochlore unit cell lattice (consisting of 65536 spins). In our approach, each step starts by generating a new candidate state for the system, as described below. Next, $S(\mathbf{q})$ is calculated using a FFT-based approach (Appendix \ref{sec:fft-interp}). 
Assuming cubic symmetry, $S(\mathbf{q})$ is evaluated for the three right permutations of the axes and then averaged. The result of the averaged FFT is  interpolated to match the experimental pixels in reciprocal space. Finally, the intensity of interpolated pixels is fitted to the corresponding experimental data by means of a linear regression. 
The linear model for the $i$-th pixel takes the form
\begin{equation}
            I_\text{low}(\mathbf{q}_i) - I_{\text{bkg}}(\mathbf{q}_i) =  c_1 \left[ S_\text{MC}(\mathbf{q}_i) - \frac{2}{3} f(q_i^2)  \right] + \epsilon_i,     
\end{equation}
where subscripts ``low'' and ``bkg'' refer to the low-temperature and background measured intensities and ``MC'', to the Monte Carlo structure factors; the term $\frac{2}{3} f(q^2)$ corresponds to the theoretical background magnetic scattering (see Appendix \ref{sec:fft-interp}), $c_1$ is a proportionality constant and $\epsilon_i$ are the residuals associated with each pixel intensity.  Although in principle the 6~K background data is assumed to exhibit negligible spin correlations, single-spin scattering is still present. The addition of a \textit{free} constant term did not improve the goodness of fit, as expected if only magnetic correlations change between 300~mK and 6~K. The value of $c_1$ was determined by the weighed linear least squares method, minimizing the normalized residual sum of squares $R = \sum_{i=1}^p (\epsilon_i / \sigma_i)^2 $,  where $\sigma_i$ is the statistical error associated to each pixel intensity.

Given that $R$ is expected to follow a $\chi^2_{p-1}$ distribution, where $p \gg 1$ is the number of data pixels, the probability distribution for $R$ can be approximated by a normal distribution with mean $\mu = p-1$ and variance $2\mu$.
Hence, the acceptance ratio for each new configuration was calculated as
	\begin{equation}
	\begin{split}
	    \alpha = \frac{P(R')}{P(R)} \simeq & \exp \left[- \frac{(R' - \mu)^2 - (R - \mu)^2}{4 \mu} \right] \\
	     = & \exp \left[\frac{R - R'}{4} \left(\frac{R+R'}{\mu} - 2 \right) \right] ,
	\end{split}
	\end{equation}
	where $R'$ and $R$ correspond to thethe new and old configurations, respectively.
Following the usual Metropolis-Hastings recipe \cite{Hastings1970}, the new configuration was accepted if $\alpha$ was larger than a pseudo-random number between $0$ and $1$ or, otherwise, rejected.
Convergence was assumed after $R$ reached a stationary value (negligible trend within fluctuations).

\begin{figure}
\centering
\begin{minipage}{0.49\columnwidth}
     \centering
     \includegraphics[width=0.95\columnwidth]{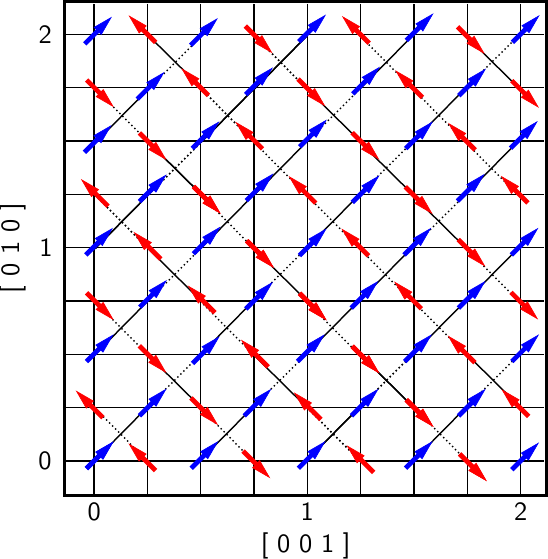}
     (a)
\end{minipage}
\begin{minipage}{0.49\columnwidth}
    \centering 
     \includegraphics[width=0.95\columnwidth]{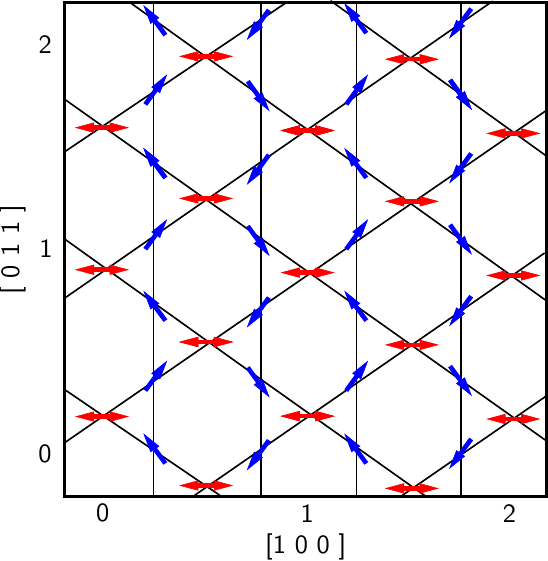}
      (b)
\end{minipage}
\caption{{\em MDG ordered state.}Two different projections of the initial ordered state: (a) cut along a $(1,0,0)$ plane and (b) cut along a $(0,1,\bar{1})$ plane.  Blue (red) spins span alternating ferromagnetic chains in $\pm [0,1,1]$ ($\pm [0,1,\bar{1}]$) directions.}
\label{fig:subs}
\end{figure}

\subsubsection{Trial configurations}

Since we are interested in low-temperature states and possible precursors to long-range order, we restricted the exploration space to suitable configurations.  Given that at low temperatures we expect to find a very low population of excitations, we restrict all ans\"atze to monopole-free configurations. As starting points, we consider two extreme cases in terms of chain order: completely disordered states, which we refer to as "monopole-free" configurations for brevity, and "spin-chain" configurations, consisting of distributions of infinite spin chains.  For each of them, a different algorithm was required in order to generate trial candidates. Intermediate order configurations are in principle possible, but they require a series of assumptions that make them impractical for a first approach. 

We used two types of initial configurations: 
\begin{enumerate}
\item The MDG state \cite{melko2001long}.
		Two projections are shown in Fig. \ref{fig:subs}. 
		Here, blue (red) spins span alternating ferromagnetic chains in $\pm [0,1,1]$ ($\pm [0,1,\bar{1}]$) directions.
		Since each tetrahedron lies in the intersection of two perpendicular chains, this configuration satisfies the two-in-two-out ice rule.
\item A random configuration. In the case of ``monopole-free'' configurations the initial configuration was created with the only constraint of the two-in-two-out rule.  For spin-chains, all spins along each $[011]$ or $[01\bar{1}]$ are made to point in the same direction, but each transect is randomly assigned its orientation. Thus, each configuration could be described as two intertwined 2-dimensional Ising states on the $(011)$ and $(01\bar{1})$ planes respectively. Predicted low-temperature states by Refs.  \onlinecite{McClarty2015,Borzi2016} would be particular cases of this general description. 
\end{enumerate}

In each case, we used different evolution mechanisms. 

\begin{enumerate}
\item {\em Spin-chain configurations.}  
Each new trial configuration was obtained by flipping all spins in one randomly chosen chain (equivalent to a single-spin flip in a 2-dimensional projected system).

\item {\em Monopole-free configurations} 
A randomly selected spin was flipped, creating a pair of opposite monopoles in two adjacent tetrahedra. 
These two monopoles were allowed to ``diffuse'' through the lattice. The diffusion of the positive (negative) monopole consisted in flipping one of the inward (outward) pointing spins of the tetrahedron hosting the monopole, picked at random.	Periodic boundary conditions were applied throughout the diffusion process.	When these two monopoles met in the same tetrahedron, they mutually annihilated, restoring the ice rule and thus ending the diffusive stage. 
 
\end{enumerate}	
	
\subsubsection{Energy calculation}

Along the optimization, configuration energy was sampled at logarithmically spaced steps.  This calculation was used for reference and did not determine the optimization process. The Hamiltonian used was the effective Hamiltonian optimized for \DTO, which includes exchange interactions up to third-nearest neighbors, discriminating between third-nearest neighbors connected through the lattice by a minimum of one or two other lattice sites, and dipolar interactions:   
\begin{displaymath}
\mathcal{H} \!=\! \!\sum_{k} \! J_k \!\!\sum_{\langle i,j \rangle_k} \!\!\mathbf{S}_i \cdot \mathbf{S}_j 
+ D r_1^3 \sum_{i,j} \!\left[ \frac{\mathbf{S}_i \!\cdot\! \mathbf{S}_j}{|\mathbf{r}_{ij}|^3} 
- \frac{3 (\mathbf{S}_i \!\cdot\! \mathbf{r}_{ij}) (\mathbf{S}_j \!\cdot\! \mathbf{r}_{ij})}{|\mathbf{r}_{ij}|^5} \right]\!
\end{displaymath}
here $k$ runs over first, second and the two types of third nearest neighbors, $J_k$ are the $k$-th neighbor interactions, $\langle i,j \rangle_k$ runs over $k$-th neighbors, $r_1$ is the nearest-neighbour distance, $\mathbf{r}_{ij}$ is the distance between spins $i$ and $j$, $\mathbf{S}_i$ is a classical spin of unit length, and the dipolar constant, $D$, is of the same order than $J_1$. We used the optimal parameters derived in \cite{Borzi2016,samarakoon2020machine} and evaluated the energy using Ewald summations to take into account the long-range interactions \cite{melko2001long}. For each nominal step, energy was averaged over ten independent realizations of the optimization algorithm. 

\begin{figure}
\centering
\includegraphics[width=\columnwidth]{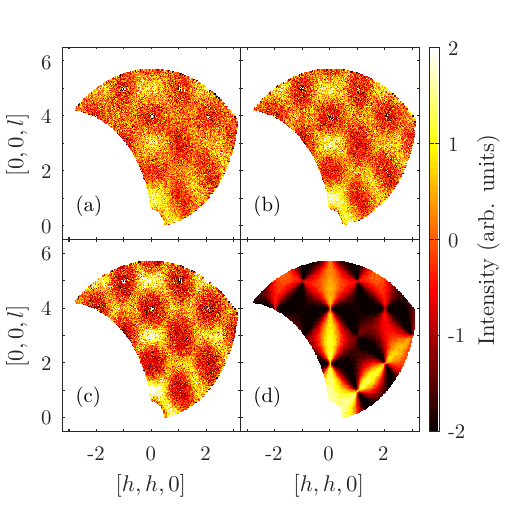}
\caption{{\em Neutron diffuse scattering of \DTO}. Measured at nominal temperatures of $800~\mK$, (a), $500~\mK$, (b), and $300~\mK$, (c), during the slow cooling shown in Fig. \ref{fig:cooling}. All panels have the same scale and had the same background subtracted (6~K). A simulated pattern corresponding to random \textit{two-in-two-out} configurations is shown in (d) for comparison.}
\label{fig:neutron_data}
\end{figure}

\section{Results and discussion}

Figure \ref{fig:neutron_data} shows in panels (a) to (c) diffuse neutron scattering measured during the slow cooldown of Figure \ref{fig:cooling} at nominal temperatures of $800~\mK$, $500~\mK$ and $300~\mK$ respectively. The waiting time of the first two temperatures amply exceeds both experimentally estimated characteristic times and can be considered to be in equilibrium. For the lowest temperature point (nominal $300~\mK$) the real temperature depends on the estimate and lies somewhere between $T_{\rm Arr}=345(10)~\mK$ and $T_{\rm VFT}=420(10)~\mK$. In all cases, the same high-temperature background (measured at $6~\K$) was subtracted.

The diffuse neutron scattering pattern seen in the experiments consists of a hexagonal mesh, matching the edges of the first Brillouin zone, with highest intensities at integer $(h,h,l)$ points with odd $h$ and even $l$. As the temperature is lowered, the diffusion pattern sharpens, without any qualitative change in its shape.  While strong scattered intensity for $\mathbf{q}$ on the border of the Brillouin zone is associated with antiferromagnetic order, its diffuse nature indicates that such order is just short-ranged even at the lowest measured temperatures. Crucially, however, this short-range order is not obtained in a random spin-ice configuration (i.e., any random arrangement satisfying the two-in-two-out rule). Such random configurations produce a distinctly different pattern, characterized by the presence of pinch-points, shown for comparison in panel (d) of this figure.

\begin{figure}
 \centering
\includegraphics[width=\columnwidth]{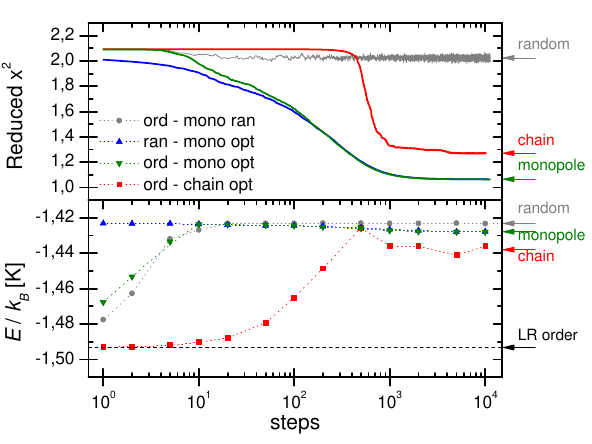}
\caption{{\em RMC analysis of experimental data.} Upper panel: evolution of the reduced $\chi^2 = R/(p - 1)$ for the different initial states and evolutions.  initially ordered (\texttt{ord - mono opt}, green)  and initially random (\texttt{ran-mono opt}, blue) configurations under monopole-free optimization, and initially ordered  (\texttt{ord-chain opt}, red) under chain-ordered optimization. Curves are averages over ten independent realizations. Gray curve and symbols  (\texttt{ord-mono ran}) correspond to an initially ordered configuration which was disordered by random monopole diffusion, without any fit to the data.
Lower panel: configuration energy along the optimization process for each case. Symbols indicate the step for evaluation, dotted lines are a guide for the eye. Arrows on the right side indicate final values for monopole-free (\texttt{monopole}) and chain-ordered (\texttt{chain}) configurations, as well as the energy of the long-range ordered initial configuration (\texttt{LR order}).
}
\label{fig:evolution}
\end{figure}

Having ruled out long-range order at the lowest measured temperature, we now turn to investigating whether the measurements reveal precursors of the low-temperature long-range ordered state --specifically, whether they contain spin chains. To address this, we perform reverse Monte Carlo (RMC) simulations on the nominal $300~\mK$ diffuse scattering data, using spin-chain and monopole-free trial configurations and using two different monopole-free initial states, random and chain ordered, as outlined in the previous section.

The upper panel of Fig. \ref{fig:evolution} shows the evolution of the reduced  $\chi^2 = R/(p-1)$ as a function of the RMC steps for the spin chain trial evolution, starting from an ordered state (red curve), and for monopole-free trial configurations starting both from ordered and random initial configurations (green and blue respectively). Fig. \ref{fig:evolution} also shows  (grey curve) the evolution of $\chi^2$ of an initially ordered state under free monopole diffusion, that is, without any optimization constraint.  After $\approx 10^3$ RMC steps the system reaches a stationary $\chi^2$ regardless of initial conditions and restrictions. Although chain-ordered configurations reproduce the experimental pattern to a good approximation, monopole-free configurations give a better fit in terms of $\chi^2$, independent of the initial conditions.

\begin{figure*}
\centering
\includegraphics[width=\linewidth]{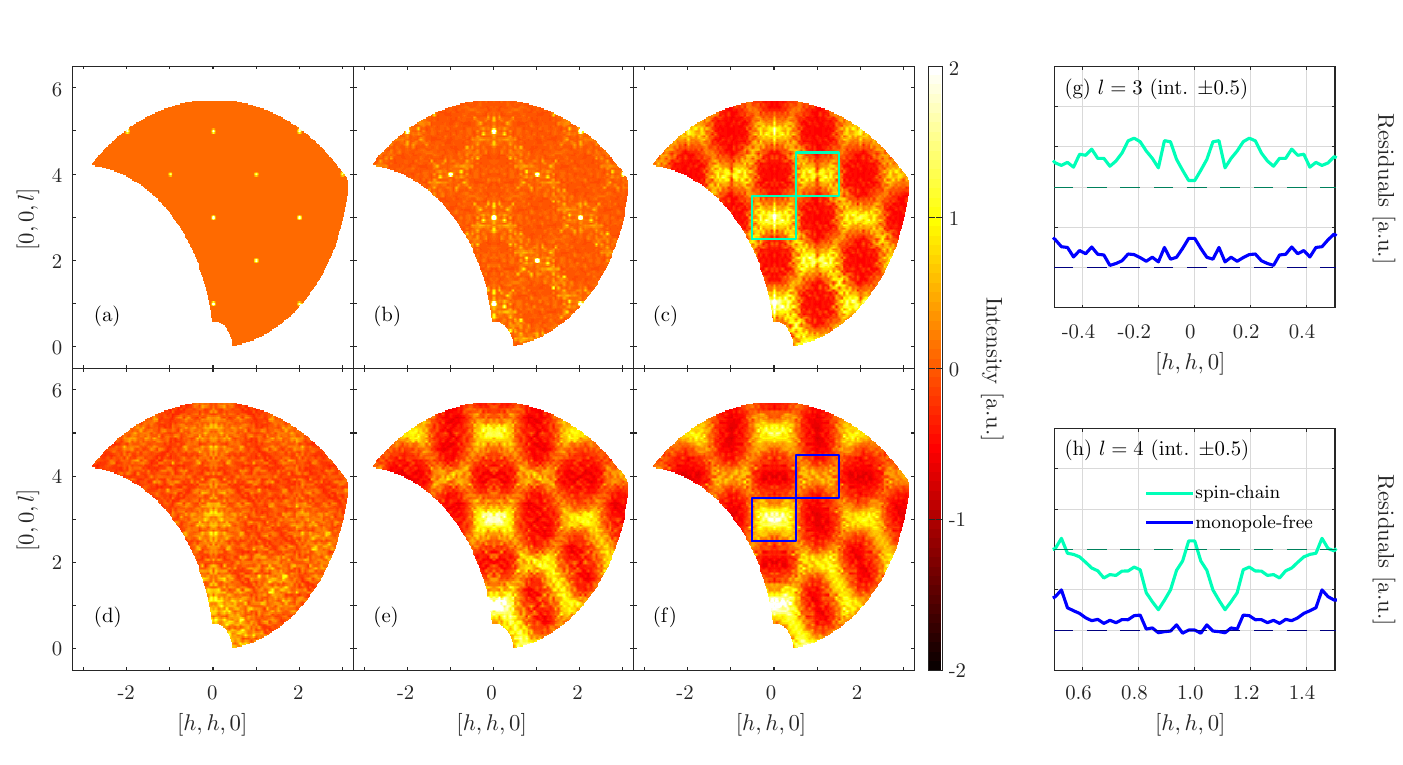}
\caption{{\em RMC diffraction patterns.} Evolution of the calculated diffraction pattern for initially-ordered spin-chain (a--c) and initially-random monopole-free (d--f) configurations, evaluated at steps 1 (a, d), 500 (b, e) and 10000 (c, f) of the reverse Monte Carlo optimization. Panels are shown in the same scale as in Fig. \ref{fig:neutron_data}. Panels (g) and (h) show the symmetrized, $l$-integrated residuals over the highlighted regions in panels (c) and (f), after averaging residuals over all available translation operations. Dashed lines indicate the baseline, as these curves were shifted for a better visualization.}
\label{fig:sq_evo}
\end{figure*}

The preference of configurations that are not based on infinite spin chains can be visually confirmed by comparing the calculated $S(\mathbf{q})$ for each case as they evolve towards the optimal configuration.  This is shown in Fig. \ref{fig:sq_evo}.  The top panels show the evolution of $S(\mathbf{q})$ starting from spin-chain ordered states, with sharp Bragg peaks at $[h,h,l]$ points with odd $h+l$, shown in panel (a). After $5\times10^2$ RMC steps, panel (b), spin-chain configurations consist of remnants of ordered regions which still yield Bragg peaks, with a diffuse halo already present in all the edges of the Brillouin zones.  When we reach step $10^4$, panel (c), the Bragg peaks have vanished and the calculated $S(\mathbf{q})$ reproduces the experimental pattern to a good approximation. The bottom panels show the evolution of the system under monopole-free configurations, starting from an initial random two-in-two-out state. Although the evolution of an initially random monopole-free state, panel (d), shares a similar mixed pattern after $5\times10^2$ steps, panel (e), it eventually reaches a state that better resembles the experimental data, panel (f). 

The most distinctive feature of spin-chain configurations is observed around points where Bragg peaks were initially located in the ordered state: the occurrence of pinch points in even-$l$ points and sharp lines (complementary \textit{stretch points}) in odd-$l$ points. These are the result of the infinite correlation along the spin chains and are absent in the less restrictive monopole-free configurations, which show a much smoother pattern at these points. 
The better agreement between calculated and experimental $S(\mathbf{q})$ for the monopole-free configurations indicates that such correlations may nucleate locally but do not develop over significantly large volumes at this temperature.  
For a cleaner comparison, residuals near these points were averaged across all translation symmetries and $h$-axis reflection and integrated along a stripe of width $l \pm 0.5$, the area shown as colored squares in panels (c) and (f). 
Details of this procedure are discussed in Appendix \ref{sec:residuals}. 
The integration centered around $l=3$ is plotted along $[h, h, 0]$  in panel (g) for both spin-chain (green) and monopole-free (blue) optimizations. 
The equivalent plots for $l=4$ are shown in panel (h).  
For the spin-chain optimization curves (green) there is a negative peak around $h=0$ for $l=3$ with a sharp counterpart positive peak for $l=4$ at $h=1$. 
These observations indicate the absence of infinite-chain correlations in the experimental data. 
Interestingly, for the monopole-free optimization, there appears to be a positive peak at $h=0$ for $l=3$ and a flat region, or even a very small dip, for $l=4$.  Despite the need of improved statistics to be conclusive, the previous fact suggests the possible presence of short-range chains in the experiments. 

The evolution of the configuration energy along the optimization process (Fig. \ref{fig:evolution}b) gives useful information in this context. We analyze the evolution of the energy for four different situations, one starting from a random spin-ice state, evolving with monopole-free constraints (in blue), and three starting from a spin-chain ordered state and evolving under increasingly unconstrained conditions: chain-ordered states (red), monopole-free states (green) and random evolution (gray). As expected, the three evolutions starting from ordered states have a significantly lower initial energy ($\simeq -1.49$~K/spin) than the random monopole-free configuration ($\simeq -1.42$~K/spin). As the RMC evolves, the initially ordered states increase monotonically in energy, but remain energetically favorable compared with the random two-in-two-out state ($\simeq -1.44$~K/spin).
In contrast, after $\sim 10^1$ steps, the breaking of spin chains in the green and gray cases takes the energy to an intermediate value similar to the random monopole-free energy. As the evolution carries on, the energy of both guided monopole-free configurations (blue and green) converge into a similar value, ($\simeq -1.43$~K/spin), slightly below the random evolution (gray) and slightly above the infinite spin-chain constrained one. This indicates not only that the experimental pattern leads to configurations with energy below the random spin-ice average, showing the development of short or medium range order, but also points to the existence of an energy barrier between this state and the one with infinite chains.

\section{Conclusions}

In spin-ice, as in many other frustrated systems, the complex energy landscape resulting from long-range dipolar interactions and dynamic constrains on excitations lead to extremely slow relaxation even in the absence of disorder \cite{hallen2022dynamical}. This makes the experimental determination of a ground state an extremely challenging problem.  In this work, we have analyzed neutron scattering data from one of the cleanest examples of spin-ice materials, isotopic \DTO sample, collected after a long cooling procedure \cite{samarakoon2022structural}.  The experiments reveal a smooth, diffuse scattering pattern that is evidently inconsistent with long-range order. Our analysis, based on reverse Monte Carlo calculations, indicates that while infinitely long chains are absent in the experimental configuration, states constructed purely by antiferromagnetic spin chains give good approximations to the data.   The main difference regarding the goodness of fit between different types of trial configurations lies in the presence of pinch points when ferromagnetic chains are forced to have infinite longitudinal correlation.  There are tantalizing signs that short-range chains may be present in the experimental data, but additional investigation is required to confirm this with certainty.

\begin{acknowledgments}
The work of D.A.T. was aided by the University of Tennessee Materials Research Science \& Engineering Center – The Center for Advanced Materials and Manufacturing – supported by the National Science Foundation under DMR No. 2309083. 
The work of M.M.B. was partly supported by grants from ANPCyT (PICT 2020 No. 372) and Universidad de Buenos Aires (UBACyT 2023 No. 20020220400234BA).
\end{acknowledgments}

\appendix

\section{Details on the calculation of $S(\textbf{q})$}
\label{sec:fft-interp}

After a collimated and quasi-monocromatic beam of neutrons crosses a lattice of spins $\mathbf{s}_n$ located at sites $\mathbf{r}_n$, the likelihood of a shift $\mathbf{q}$ in neutrons' momentum is proportional to an effective structure factor
\begin{equation}
    S(\mathbf{q}) = \frac{f(q^2)}{N} ~ \left|
    \sum_{n} \textbf{s}_n^{\perp}(\mathbf{q}) ~  e^{i \mathbf{q} \cdot \mathbf{r}_n}
    \right|^2,
    \label{eq:neutronSq}
\end{equation}
where $f(q^2)$ is the atomic form factor and, introducing the pseudo-spin $\sigma_n = \pm 1$, 
\begin{equation}
    \textbf{s}_n^{\perp}  (\mathbf{q}) = \textbf{s}_n -
     \mathbf{q} \frac {\textbf{s}_n \cdot \mathbf{q}} {q^2} = \sigma_{n} ~ \hat{s}_n^{\perp}(\hat{\mathbf{q}})
     \label{eq:normspinproj}
\end{equation}
is the projection of spin $\textbf{s}_n$ normal to $\mathbf{q}$. 
 
At its core, evaluating $S(\textbf{q})$ for a given spin configuration involves calculating its Fourier transform.
The computational cost of this process can be greatly reduced by taking advantage of the fast Fourier transform (FFT) algorithm.
This can be done effectively after rearranging the terms in Eq. \ref{eq:neutronSq}.
The spins are labeled hierarchically, $\mathbf{r}_n = a_0 ( \mathbf{R}_{ijk} + \mathbf{R}'_{a} + \mathbf{R}''_{b})$, where  $\mathbf{R}_{ijk} = i \hat{x} + j \hat{y} + k \hat{z}$ marks the position of the unit cell,  $\mathbf{R}'_{a} \in \{\mathbf{0}, (\hat{x} + \hat{y})/2, (\hat{x} + \hat{z})/2, (\hat{y} + \hat{z})/2\} $, labels the tetrahedra within the cell and $\mathbf{R}''_{b} \in \{(\hat{x} + \hat{y} + \hat{z})/4, \hat{y}/4, \hat{z}/4, \hat{x}/4\} $ labels the positions within each tetrahedron. The exponential in Eq. \ref{eq:neutronSq} can then be factored as
\begin{equation}
    e^{i \mathbf{q} \cdot \mathbf{r}_n} = 
    e^{i \mathbf{Q} \cdot\mathbf{R}_{ijk}} ~ e^{i \mathbf{Q} \cdot\mathbf{R}'_{a}} ~ e^{i \mathbf{Q} \cdot\mathbf{R}''_{b}},
\end{equation}
with $\mathbf{Q} = a_0 ~ \mathbf{q}$, the adimensionalized wave vector.  
Since the local $[1,1,1]$ direction depends only on the spin coordinate $b$,it is convenient to arrange the sum in Eq. \ref{eq:neutronSq} as
\begin{equation}
\begin{split}
   \mathbf{G}(\mathbf{Q}) \overset{\text{\scriptsize def}}{=}&  \sum_{n} \textbf{s}_n^{\perp}(\hat{\mathbf{q}}) ~  e^{i \mathbf{q} \cdot \mathbf{r}_n} \\
     = & \sum_{b} \hat{s}_b^{\perp}(\mathbf{Q}) ~  e^{i \mathbf{Q} \cdot\mathbf{R}''_{b}}
    \sum_{a} e^{i \mathbf{Q} \cdot\mathbf{R}'_{a}} ~ F_{ab}(\mathbf{Q}),
\end{split}
\label{eq:G_q}
\end{equation}
where $F_{ab}(\mathbf{Q})$ is the Fourier transform of the cubic sublattice formed by all spins in sites with coordinates $a,b$.
By taking into account that experimental data is restricted to the $(h,h,l)$ plane, 
\begin{equation}	
\begin{split}
F_{ab}(h,l) = & \sum_{i,j,k} \sigma_{ijkab}~ e^{2i\pi [(i+j) h + kl]} \\
= & \sum_{k} e^{2i\pi k l}  \sum_{m} e^{2i\pi m h} \sum_{j} \sigma_{(m-j)jkab} 
\end{split}
\end{equation}
and can be evaluated as a two-dimensional discrete Fourier transform on the pseudospin sums along the $[-1,1,0]$ direction.
Hence, the FFT algorithm can be applied to reduce calculation complexity from $\mathcal{O}(L^5)$ to $\mathcal{O}(L^3 \log^2 L)$, enabling the study of large systems, as all other factors in Eq. \ref{eq:G_q} can be precalculated and reused for a given lattice size.

Lastly, we show that the sum in Eq. \ref{eq:neutronSq} can also be split into a fixed term plus a modulation introduced by specific spin configurations. 
Expanding 
\begin{equation}
\begin{split}
|\mathbf{G}(\mathbf{Q})|^2 =& \mathbf{G}^*(\mathbf{Q}) \cdot \mathbf{G}(\mathbf{Q}) \\ =& \sum_{m,n} \textbf{s}_m^{\perp}(\mathbf{q}) \cdot \textbf{s}_n^{\perp}(\mathbf{q}) ~  e^{i \mathbf{q} \cdot (\mathbf{r}_m - \mathbf{r}_n)}  
 \end{split}
\end{equation}
and grouping all terms with $m=n$, we obtain
\begin{equation}
 \begin{split}
|\mathbf{G}(\mathbf{Q})|^2 =& \sum_{n} |\textbf{s}_n^{\perp} (\mathbf{q})|^2 \\ +& \sum_{m \neq n} \textbf{s}_m^{\perp}(\mathbf{q}) \cdot \textbf{s}_n^{\perp}(\mathbf{q}) ~  e^{i \mathbf{q} \cdot (\mathbf{r}_m - \mathbf{r}_n)}.    
 \end{split}
 \label{eq:spinsumgrouping}
\end{equation}
Since  $\textbf{s}_n^{\perp}  (\mathbf{q}) = \sigma_{n} ~ \hat{s}_b^{\perp}(\mathbf{Q})$, $\sigma_{n}^2 = 1$, and the four sites labeled by $b$ are equally represented in the lattice, 
\begin{equation}
	 \frac{1}{N} \sum_{n} |\textbf{s}_n^{\perp} (\mathbf{q})|^2 =  \frac{1}{4} \sum_{b} \left| \hat{s}_b^{\perp}(\mathbf{Q}) \right|^2 = \frac{2}{3}.
	\end{equation}
Therefore, even if spins are uncorrelated (thus, the expectation value of the second sum in Eq. \ref{eq:spinsumgrouping} is null), the structure factor takes the finite value
\begin{equation}
    S_{\text{unc}}(\mathbf{Q}) =  \frac{2}{3} ~ {f(Q^2)} .
\end{equation}
 
\begin{figure}[t!]
\centering
\includegraphics[width=0.9\columnwidth]{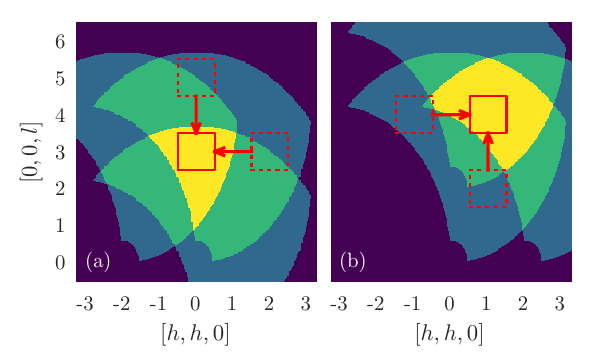}
\caption{Translations applied to average over odd- (a) and even-$l$ (b) regions. The shaded areas show how data limits overlapped across translations}
\label{fig:overlap}
\end{figure}

\section{Residuals averaging and integration}
\label{sec:residuals}
\begin{figure*}[bt]
\centering
\includegraphics[width=\linewidth]{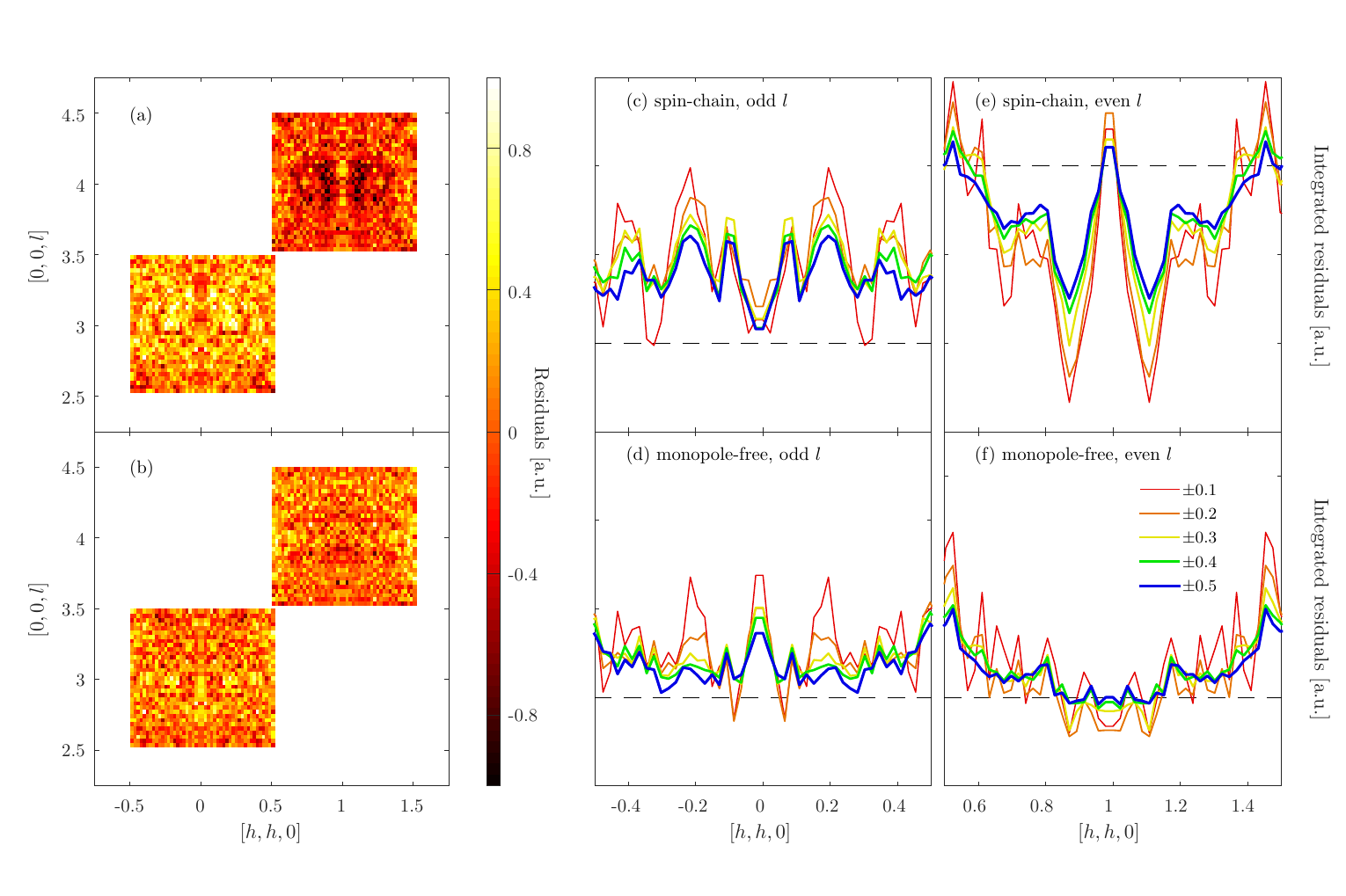}
\caption{Translation-averaged residuals, calculated as the difference between the averaged data and the optimized diffraction patterns, for (a) spin chains and (b) monopole-free optimizations. Residuals were integrated along a stripe centered around $l=3$ (c,d) and $l=4$ (e,f).  The integration width $\Delta l$ for each curve is shown in the figure legend. In each case, residuals for chain (c,e) and monopole-free optimizations (d,f) are shown.}
\label{fig:residuals}
\end{figure*}

By simple inspection of both experimental and simulated $S(\mathbf{Q})$, local maxima at even $h+l$ points can be classified into two kinds, corresponding to odd and even $l$ values.
Hence, for a better signal-to-noise ratio in the analysis of the residuals (i.e., the difference between the averaged data and the optimized diffraction patterns), these were averaged across translation symmetries preserving $l$ parity.
For each parity, three regions of size $\pm 0.5$ in each direction around maxima centers were available (within detector limits), corresponding to the translations shown in Fig. \ref{fig:overlap}.
After averaging across analogous regions, residuals were symmetrized with respect to $h \rightarrow -h$ reflections to obtain the patterns shown Figures \ref{fig:residuals}(a,b). 
Lastly, we performed line-cuts along $[h,h,0]$ at fixed $l$ and integrated along a stripe of varying width ($\Delta l$).  
Figure \ref{fig:residuals} shows this analysis for $l=3$ (c,d) and $l=4$ (e,f).  
The integration shows that the main features seen around $h=0$ for $l=3$ and $h=1$ for $l=4$ are robust with respect to the choice of $\Delta l$. 

\bibliographystyle{apsrev4-2}.
\bibliography{postdoc}
	
\end{document}